%% file: main.tex
\documentclass[journal ,12]{IEEEtran}
\usepackage[pdftex]{graphicx}
\graphicspath{{../pdf/}{../jpeg/}}
\DeclareGraphicsExtensions{.pdf,.jpeg,.png}
\usepackage[cmex10]{amsmath}
\usepackage{amsthm}
\usepackage{graphicx,color}
\usepackage[dvipsnames]{xcolor}
\usepackage{graphicx}
\usepackage{mathtools}
\usepackage{amssymb,color}
\allowdisplaybreaks
\usepackage{subcaption}
\usepackage{cite}
\usepackage{blindtext}
\usepackage{tcolorbox}
\usepackage{colortbl}
\usepackage{booktabs}
\definecolor{Gray}{gray}{0.9}
\usepackage{xcolor}
\usepackage{lipsum}
\usepackage[font={small}]{caption}
\usepackage{algorithm}
\usepackage{algorithmic}
\usepackage{array} 
\usepackage{makecell}
\usepackage{tabularx}
\usepackage{bm}
\usepackage{siunitx, soul}
\usepackage{pgfplots}
\pgfplotsset{compat=newest}
\usetikzlibrary{plotmarks}
\usetikzlibrary{arrows.meta}
\usepgfplotslibrary {patchplots}
\usepackage{grffile}
\usepackage[font=scriptsize,labelfont=bf]{caption}
\usepackage{pifont}

\newcommand{\bc}{\text{BackCom}\xspace}

\usepackage{enumitem}
\usepackage{multirow}

\makeatother
\bstctlcite{IEEEexample:BSTcontrol}

\definecolor{LightGray}{gray}{0.9}
\definecolor{MediumGray}{gray}{0.5}
\definecolor{DarkGray}{gray}{0.2}

\begin{document}
\bstctlcite{IEEEexample:BSTcontrol}

\title{Sensing and Backscatter Communication Integration: Realizing  Efficiency in Wireless Systems for IoT}

\author{Shayan Zargari, Diluka Galappaththige,  and Chintha Tellambura
\thanks{S. Zargari, D. Galappaththige, and C. Tellambura with the Department of Electrical and Computer Engineering, University of Alberta, Edmonton, AB, T6G 1H9, Canada (e-mail: \{zargari, diluka.lg, ct4\}@ualberta.ca).} \vspace{-5mm}}

\maketitle

\begin{abstract} 
In an era driven by the Internet of Things (IoT) and rapid wireless communication advances, the synergy between sensing and backscatter communication (BackCom) has emerged as a frontier of research. This paper delves deep into the integration of sensing technologies with BackCom, a burgeoning field with significant implications for energy-efficient wireless systems. By tracing the historical developments and principles of BackCom, we establish the foundational understanding crucial for integrating advanced sensing methodologies. Our study adopts a mixed-method approach, combining quantitative analyses of system performances with qualitative assessments of various integration techniques. Furthermore, these integrated systems showcase enhanced adaptability in dynamic environments, a pivotal attribute for future IoT applications. These findings hold profound implications for industries focusing on smart technologies, as they underscore the potential for achieving both sustainable and efficient communication. Our research distinguishes itself by not only highlighting the benefits but also addressing the challenges faced in this integration, providing a comprehensive overview of the topic.
\end{abstract}


\IEEEpeerreviewmaketitle

\section{Introduction}
The dawning of the digital epoch is epitomized by the Internet of Things (IoT)—an era where the tangible intertwines seamlessly with the virtual, weaving a cohesive narrative of inter-connectedness and data exchange \cite{Huawei_ambient}. Propelling this monumental shift is the remarkable stride in wireless communication and the miniaturization of both computing and sensing modules. At the very heart of this transformation lies massive machine-type communication (mMTC), which orchestrates a harmonious dialogue between devices and central mMTC servers, ensuring a consistent modus operandi across an expansive spectrum of applications \cite{3GPP2017study}.

However, the momentum of this digital revolution, anchored by mMTC, encounters certain roadblocks. The energy needed to power devices, the cost of deploying them in large numbers, and the limited ways they can communicate all present unique challenges \cite{3GPP2017study}. Sensing, the very essence of myriad IoT applications, plays a pivotal role, ranging from discerning the whispers of our environment to tracking the nuances of biological entities. Simultaneously,  the advent of backscatter communication (BackCom) has emerged as a promising panacea. In particular, ambient BackCom (AmBC) sidesteps energy-intensive traditional communication by leveraging pre-existing ambient signals and modulating them for data transmission \cite{HoangBook2020}.

\begin{table*}[t]
\centering
\captionsetup{font=small,labelfont={color=DarkGray,bf},textfont={color=DarkGray}}
\caption{Features and Challenges of AmBC Systems}
\begin{tabularx}{\linewidth}{>{\raggedright\arraybackslash}p{2.5cm} 
                              >{\raggedright\arraybackslash}p{4.3cm} 
                              >{\raggedright\arraybackslash}p{4.3cm} 
                              >{\raggedright\arraybackslash}p{5.4cm}}
\toprule[1.5pt]
\rowcolor{LightGray}
\textbf{\color{DarkGray} Features} & \textbf{\color{DarkGray} Challenges} & \textbf{\color{DarkGray} Solutions} & \textbf{\color{DarkGray} Advantages} \\
\midrule[1pt]
Energy Efficiency & Limited range and low data rate due to passive reflection and modulation of ambient RF signals. & Enhancement of EH techniques and optimization of modulation schemes. Utilization of surrounding signals broadcast from ambient RF sources. & Battery-free operation within a power range of \qty{1}{\uW} to \qty{100}{\uW}, reducing the environmental impact. Enables communication without requiring active RF transmission. \\
\midrule
Scalability & Interference from other RF devices and varying signal availability affect reliability. & Advanced signal processing and filtering techniques to mitigate interference and enhance signal quality.  & Supports up to \num{100} connections connections per \qty{}{m^3}, preventing network overload and enhancing adaptability in diverse environments.   \\
\midrule
Cost-effectiveness & Complexity and cost of designing specialized hardware and software for AmBC. & Development of cost-effective and efficient alternatives to traditional radio components. & Low-cost devices ($0.01\sim0.5$  each) with extended lifespan exceeding \num{10} years, suitable for mass deployment. Minimization of RF components at backscatter devices. \\
\midrule
Ubiquity & Ethical and legal concerns regarding unauthorized use of spectrum resources and security vulnerabilities. & Utilization of pervasive ambient signals with enhanced security protocols and adherence to legal frameworks. & Operation in diverse environments, ranging from coverage of $100-200$ m for industrial areas and $2-5$ m for wearables. Does not require a dedicated frequency spectrum which is scarce and expensive. \\
\bottomrule[1.5pt]
\end{tabularx}
\label{tab:ambc_overview_revised}
\end{table*}

The amalgamation of AmBC into the IoT landscape promises significant transformations, as highlighted in Table \ref{tab:ambc_overview_revised}. However, this fusion extends beyond this initial integration. With the evolution of integrated sensing and communication (ISAC) systems, the convergence of communication capabilities with sensing prowess is underway \cite{Behravan2023, Zhang2021, Wang2022, Tong2023}. When this fusion intersects with backscatter technology, it gives rise to what we call integrated sensing and BackCom (ISABC) \cite{Diluka2023}. This convergence effectively elevates passive elements to become proactive data communicators, enhancing both the granularity of communication and the precision of sensory data. However, it also introduces the challenge of complicated decoding processes.

This paper aims to clarify the relationship between AmBC and ISAC for both scholars and industry experts. We initiate with a comprehensive examination of prevalent AmBC and ISAC architectures, transitioning to the intricacies of ISABC, analyzing its merits, confronting inherent challenges, and proffering possible solutions. Our discussion leads to mapping out possible future paths, creating a comprehensive outlook for the intersection of IoT, AmBC, and ISABC.

\section{Integration of Sensing with BackCom}
The fusion of sensing capabilities with BackCom has given rise to a novel paradigm where devices can not only perceive their environment but also communicate their findings in a highly energy-efficient manner. The integration of these features is especially pertinent in the IoT field, where devices frequently have to function in low-energy settings for a long duration.

\subsection{Fundamentals of BackCom}
BackCom is an ultra-low-power communication paradigm. Unlike traditional networks, Bakcom uses tags that do not generate their own RF signal. Instead, they use passive backscatter modulation to reflect the incoming RF signals for data transmission. Thus, BackCom achieves higher energy efficiency, superior spectrum utilization, and cost-effectiveness, making it crucial for future wireless networks \cite{Diluka2022, Rezaei2023Coding}.

To visualize the foundational components of a simple BackCom network, we consider a network of a tag, a reader, and an emitter. The tag, equipped with a low-power integrated circuit (IC), primarily reflects incoming RF signals to transmit data. It also harvests energy from these signals for its operations, leading to two design types: (i) passive tags without energy storage and (ii) semi-passive tags with limited energy storage \cite{Diluka2022}.

The reader is a more robust component equipped with its own power supply and full-fledged RF components, empowering it to execute intricate RF functions, including demodulation and decoding, essential to retrieve data from tags. The reader can manifest in various forms, from mobile phones to wireless fidelity (Wi-Fi) access points (APs). Finally, the system's emitter, also known as the RF source, is responsible for producing the RF signal that the tags reflect. An emitter can be either a dedicated beacon signal producer or an ambient transmitter, such as a television (TV) tower or a cellular base station (BS).

\subsubsection{Configurations of BackCom}
BackCom has several configurations, each tailored for specific application scenarios \cite{Diluka2022, Rezaei2023Coding}. Here is a detailed overview:

\begin{itemize}
    \item \textbf{Monostatic BackCom (MoBC):} In MoBC, the reader assumes a dual role by transmitting the RF signal and decoding the signals reflected by the tag. However, this setup presents its own set of challenges. As signals travel from the reader to the tag and back again, they encounter what is known as "dyadic" or round-trip path loss. Specifically, when the tag and reader are significantly spaced apart, the initial RF signal received by the tag may be too weak to power it adequately. Additionally, the reflected signal on its return journey to the reader may also be too faint for accurate detection. Given these inherent limitations, MoBC naturally finds its most suitable applications in short-range scenarios, such as radio-frequency identification (RFID).

    \item \textbf{Bistatic BackCom (BiBC):} Departing from MoBC's setup, BiBC introduces two separate entities; a dedicated RF carrier emitter and a reader. This division tackles MoBC's twin challenges head-on. Firstly, it expands the limited range of the forward link. Secondly, it tackles the challenging round-trip path loss of MoBC. BiBC also provides a safeguard against the ``doubly near-far problem." 

    \item \textbf{Ambient BackCom (AmBC):} Instead of deploying new RF emitters, it recycles signals from existing ambient ones. Consequently,  the overhead of deploying and upkeeping dedicated RF emitters evaporates, leading to substantial cost and energy savings. Furthermore, the valuable spectrum can be optimized as additional allocations become unnecessary. However, the path of innovation is seldom without its potholes. The unpredictable nature of ambient RF signals may interfere directly with readers. Moreover, designing an efficient AmBC system is like trying to hit a moving tag, especially when characteristics of ambient RF emitters, like power outputs and locations, are unpredictable and beyond control.
    
\end{itemize}

\begin{figure*}[!t]
    \centering 
    \def\svgwidth{510pt} 
    \fontsize{9}{9}\selectfont 
    \graphicspath{{Figures/}}
    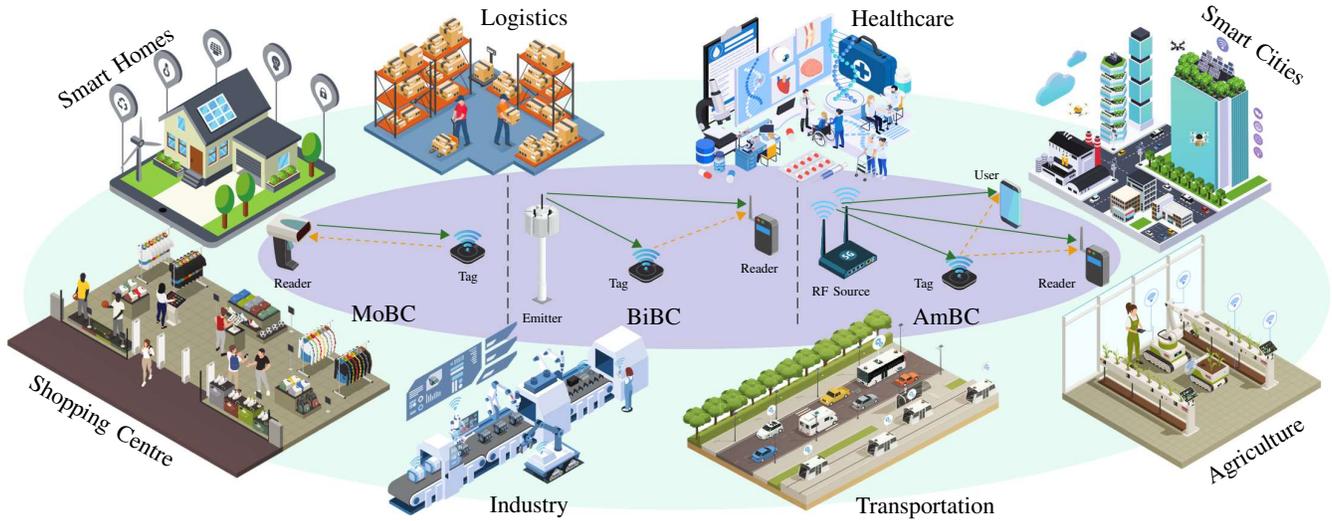  
    \vspace{1mm}
    \caption{\bc configurations and applications. }  \label{fig_BCApplications}
\end{figure*}

Fig. \ref{fig_BCApplications} depicts the BackCom configurations and some of its applications. Readers interested in the fundamentals of BackCom and its applicability may refer to \cite{Diluka2022, Rezaei2023Coding} and the references therein.

\subsection{Integrated Sensing and Communication}
ISAC epitomizes a transformative stride in the realm of wireless technology, melding the capabilities of communication and sensing into one unified framework \cite{Behravan2023, Zhang2021, Wang2022, Tong2023}. This interdisciplinary domain does not just enable devices to communicate; it allows them to actively perceive and interpret their surroundings. The ISAC aims to provide dual functionalities. While traditional communication systems have been exclusively about transmitting and receiving data, ISAC broadens this by using the very same communication signals to sense the environment. This not only enhances the potential applications but also leads to an efficient use of resources. By using the same spectrum and hardware for both communication and sensing, systems can minimize redundancy and save on critical resources \cite{Wei2022, Ouyang2023InfFramework}.

The essence of ISAC revolves around certain fundamental components. The transceiver architecture, which is the core of any communication device should be meticulously designed to seamlessly support both communication and sensing functionalities without one interfering with the other. Waveforms, the signals used in communications, need to be tailored for ISAC, ensuring they are versatile enough to cater to the dual requirements of data transmission and environmental sensing. To extract meaningful information from these waveforms, cutting-edge signal processing algorithms are paramount. These algorithms must be adept at gleaning sensing data from communication signals and vice-versa.

Given its inherent nature, ISAC places a significant emphasis on spatial information-related sensing parameters \cite{Wei2022, Ouyang2023InfFramework}. These include the direction of arrival (DoA) which ascertains from which direction a signal is coming, the signal propagation time delay which gauges the time taken for a signal to travel between a transmitter and receiver, and the Doppler frequency which gives insight into the frequency changes due to relative motion between the source and the receiver. For applications integrated with radar sensing, parameters such as the position and velocity of moving objects become incredibly pertinent. One of ISAC's key inspirations is its potential integration with radar functionalities, which are paramount in applications requiring the identification of spatial parameters of objects \cite{Wei2022, Ouyang2023InfFramework}. In an ISAC system, the essence of sensing is derived from the wireless channel itself, thereby emphasizing spatial parameters. This not only aids in primary sensing objectives but also sets the stage for intricate tasks such as imaging, recognition, and classification \cite{Wei2022, Ouyang2023InfFramework}. 

The present direction of ISAC indicates a promising future. The formation of specialized groups such as IEEE 802.11bf, coupled with the dedication of the wireless local area network (WLAN) sensing topic interest group, highlights the growing enthusiasm in this field \cite{Meneghello2023}. Additionally, the surge in academic investigations, from transceiver designs to waveforms, further underscores the field's burgeoning momentum.

\subsection{Integrated Sensing and BackCom}
To reap the benefits of both ISAC and \bc, ISABC replaces the traditional sensing/radar targets in ISAC systems with backscatter tags \cite{Diluka2023}. Thus, conventional ISAC and innovative ISABC systems differ in several key aspects. In particular, the backscatter tags serve as the sensing targets, conveying additional data to the user/reader while providing sensing information to the BS. This is essential in applications such as smart homes, warehousing, and more, where these tags can be used to learn and map the environment. ISABC, on the other hand, utilizes the unintentionally received backscattered signals at the BS for sensing purposes, needing no additional RF resources, hardware cost, or modifications to backscatter tags. Additionally, by utilizing both sensing and backscatter data, ISABC systems considerably improve communication and sensing performance; nevertheless, the complexity of the user/reader may rise due to the requirement for successive interference cancellation (SIC)-based decoding.

\begin{figure*}[!t]
    \centering 
    \def\svgwidth{500pt} 
    \fontsize{9}{9}\selectfont 
    \graphicspath{{Figures/}}
    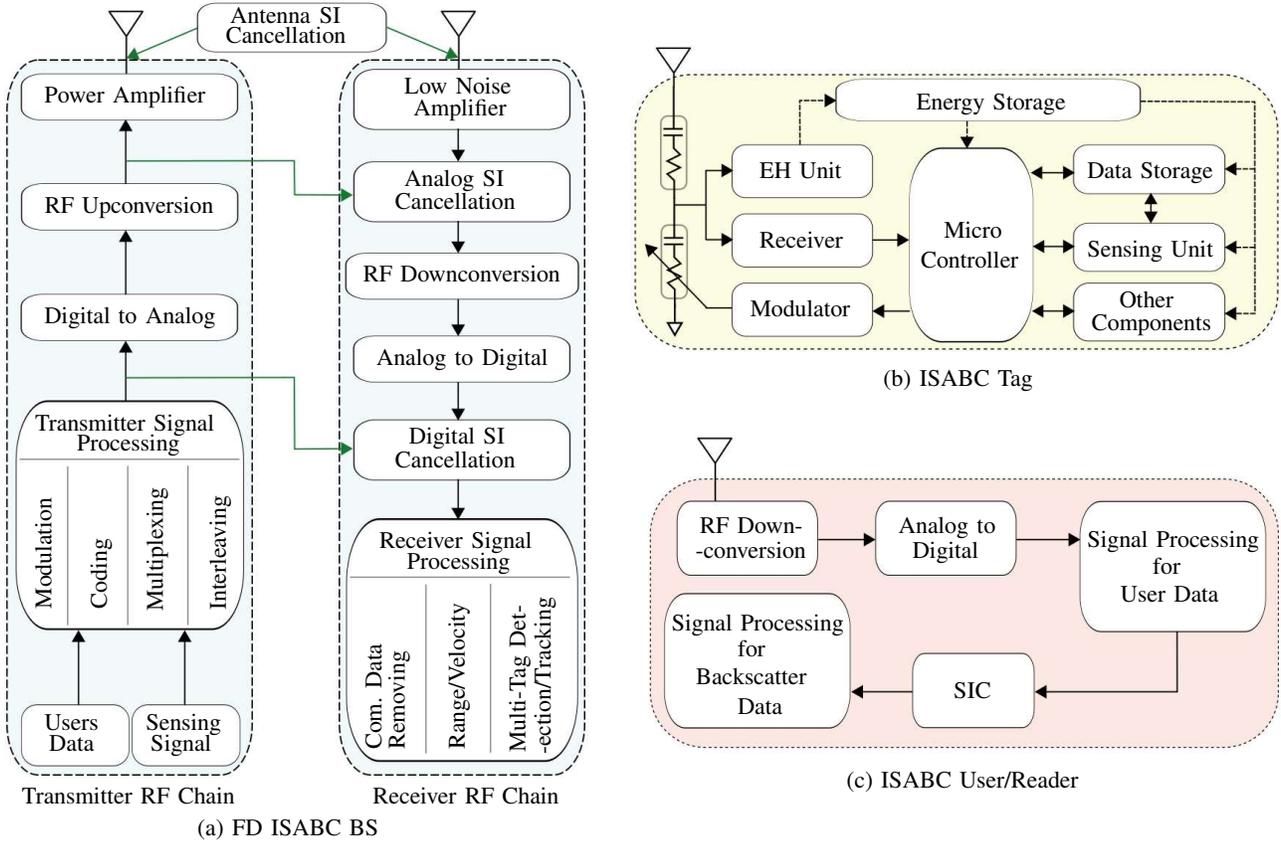  
    \vspace{1mm}
    \caption{Circuitry and signal processing of an FD ISABC BS, an ISABC tag, and an ISABC user/reader.}  \label{fig_SigProDiagram}
\end{figure*}

Building upon these foundational frameworks, we present an advanced concept termed ISABC. This concept innovatively departs from the standard sensing/radar targets intrinsic to ISAC systems and introduces the use of a passive tag. To delineate the contrasts between ISAC and ISABC systems, we highlight the following critical distinctions:

\begin{itemize}
\item \textbf{Backscatter tag utilization:} The ISABC configuration makes a notable pivot by integrating a backscatter tag. This is in contrast to the ISAC system which relies on a designated target for its sensing operations. In ISABC, the tag does not merely act as a sensing target. Instead, it assumes a dynamic role by actively supplying the BS with detailed sensing data. Simultaneously, this tag serves as a relay for auxiliary data to end users. For instance, this tag, which could also be a sensor, is capable of conveying ambient information to users, enriching them with insights into environmental parameters such as range, velocity, and orientation.

\item \textbf{Data integration:} Another distinguishing hallmark of ISABC systems is their adeptness at merging sensing and backscatter data. The synergy resulting from this integration signifies a monumental leap in terms of communication throughput and sensing precision. While ISAC systems do not provide additional data to the user, ISABC fills this gap by ensuring users receive auxiliary data. However, this advancement comes at the cost of heightened computational requirements on the user's side. The ISABC system demands advanced decoding algorithms which are propelled by SIC. Contrarily, the ISAC system employs more conventional decoding techniques.

\item \textbf{Efficiency and performance metrics:} Delving deeper into the metrics of efficiency, the ISABC system excels in spectral efficiency with a high rating as compared to the moderate spectral efficiency seen in ISAC. This likely contributes to ISABC's potential in channel training and the ensuing enhancement of spectrum efficiency by curtailing training overhead. Nevertheless, this heightened spectral efficiency in ISABC is compared with a medium latency, which is a step up from the low latency characteristic of ISAC. Furthermore, while both systems exhibit scalability and allocate power at the BS, ISABC's energy efficiency is marked low as opposed to the high energy efficiency seen in ISAC.
\end{itemize}

In summary, this pioneering concept paves the way for leveraging passive tags in both environmental monitoring and wireless communication networks \cite{Diluka2023}. Figure \ref{fig_SigProDiagram} illustrates the schematic design and signal processing flow of the proposed ISABC system with a full-duplex (FD) base station (BS) in \cite{Diluka2023}. Notably, it emphasizes the self-interference (SI) cancellation process and showcases the utilization of backscattered signals for sensing at the BS, energy harvesting (EH), modulation processing at the tag, and data decoding at the user/reader with successive interference cancellation (SIC).

\section{Passive Tags/Sensors in ISABC}
A tag consists of several components, chiefly an analog front-end. This integrates an energy harvesting (EH) system and a transceiver crucial for signal modulation and demodulation. The baseband processing unit manages data encoding/decoding and interfaces with memory or sensors. Some tags incorporate environmental sensors, essential for electric grid monitoring, livestock telemetry, and manufacturing diagnostics. Without traditional batteries, these tags rely heavily on an EH module to capture energy from ambient RF signals, while the processor and transceiver are designed for minimal power usage. The design of these tags is crucial to the ISABC framework, and current cutting-edge research aims to address design hurdles and boost efficiency.

\begin{table*}[t]
\centering
\captionsetup{font=small,labelfont={color=DarkGray,bf},textfont={color=DarkGray}}
\caption{An overview of challenges in ISABC}
\label{tab:challenges}
\begin{tabularx}{\linewidth}{>{\raggedright\arraybackslash}p{3cm} 
                              >{\raggedright\arraybackslash}p{6cm} 
                              >{\raggedright\arraybackslash}p{3.5cm} 
                              >{\raggedright\arraybackslash}p{4cm}}
\toprule[1.5pt]
\rowcolor{LightGray}
\textbf{\color{DarkGray} Categories} &\hspace{3mm}  \textbf{\color{DarkGray} Challenges}  & \textbf{\color{DarkGray} Implications} & \textbf{\color{DarkGray} Solutions}  \\
\midrule[1pt]
Environmental Factors and RF Constraints & 
\vspace{-2mm} 
\begin{itemize}
    \item Unknown ambient RF source: Uncertainty in the surrounding RF environment.
    \item Low backscatter signal strength: Challenges in achieving a strong communication signal.
    \item Ambient interference: Issues with external RF noise.
    \item Multi-user interference: Clashes when multiple users transmit simultaneously.
    \item Distance and Path Loss: Signal degradation over distance.
    \item Channel estimation \& coding: Predicting and adapting to channel conditions.
\end{itemize}
& Reduces reliability and efficiency of communication, potential data loss, and misinterpretation. & Adaptive signal processing, enhanced antenna designs, spectrum sensing for ambient RF sources, interference mitigation algorithms, and robust channel estimation techniques. \\
\midrule
Hardware and System Integration Limitations &
\vspace{-2mm}
\begin{itemize}
    \item Hardware Constraints: Limitations due to size, power, and cost.
    \item Sensing vs. Communication: Trade-offs between data collection and transmission.
    \item Integration with Legacy Systems: Challenges in compatibility with older systems.
\end{itemize}
& Affects device compatibility, limits data throughput, and increases energy consumption. & Modular hardware design, hybrid systems, backward compatibility strategies, power-aware hardware components, and middleware solutions for legacy integration. \\
\midrule
Standardization, Security, and Privacy & 
\vspace{-2mm}
\begin{itemize}
    \item Standardization and Protocol Development: Need for universal standards.
    \item Security and Privacy Concerns: Potential threats to user data and system integrity.
\end{itemize}
& Can lead to compatibility issues, security breaches, and loss of user trust. & Collaborative standardization efforts, advanced encryption, two-factor authentication, privacy-preserving algorithms, and security audits. \\
\midrule
Scalability and Network Dynamics & 
\vspace{-2mm}
\begin{itemize}
    \item Rapid expansion of the IoT ecosystem: Increased devices lead to congestion.
    \item Dynamic Network Topologies: Changing device connections and data paths.
\end{itemize}
& Alters network performance, potential bottlenecks, and degraded user experience. & Hierarchical network topologies, machine learning (ML) for adaptive scheduling, dynamic routing protocols, edge computing for local data processing, and network slicing for efficient resource allocation. \\
\midrule
Energy Efficiency and Sustainability & 
\vspace{-2mm}
\begin{itemize}
    \item Energy consumption: High power usage in continuous communication.
    \item Dependence on non-renewable energy: Environmental concerns.
\end{itemize}
& Shortening device lifetime increases operational costs and environmental concerns. & EH techniques, low-power communication protocols, and renewable energy integrations. \\
\bottomrule[1.5pt]
\end{tabularx}
\end{table*}
\subsubsection{Antenna}
BackCom devices operate over a range of frequencies, influenced by transmission protocols, regulations, and applications \cite{Diluka2022}. For instance, RFID systems span from the low-frequency band (\qty{125}{\MHz} to \qty{134}{\MHz}) to the super high frequency (SHF) band (\qty{2.4}{\GHz} to \qty{5.875}{\GHz}). Notably, ultrahigh frequency
(UHF) tags (\qty{860}{\MHz} to \qty{960}{\MHz}) are favored due to their compact antennas like microstrip, dipole, and planar designs.

Higher operating frequencies, especially in the SHF band, present distinct advantages for passive circuits. At these frequencies, antennas, such as the half-wave dipole, become smaller: \qty{16}{\cm} at \qty{915}{\MHz}, \qty{6}{\cm} at \qty{2.45}{\GHz}, and \qty{2.5}{\cm} at \qty{5.79}{\GHz}. Consequently, tags can integrate multiple antennas, leveraging advanced techniques like beamforming and space-time block coding to offset fading and enhance communication \cite{Diluka2022, Rezaei2023Coding}.

\subsubsection{Backscatter modulation}
Modulation serves to map data onto an ambient RF carrier. Unlike active radios, which generate their own carrier signal, tags modulate by reflecting external RF signals \cite{Rezaei2023Coding}. This reflection process involves the tag switching between various impedance values, giving rise to distinct reflection coefficients \cite{Rezaei2023Coding}. To create a general multi-level ($\tilde{M}$-ary) modulation scheme, the tag will use $\tilde{M}$ load impedance values. The reflection coefficient at time $k$ can be defined as
\begin{eqnarray}\label{eqn::reflection}
  \Gamma_m(k) = \begin{cases}\frac{Z_m(k) - Z^*_a}{Z_m(k) + Z_a} = \sqrt{\beta} b_m(k), \quad &  m = 1, \ldots, \tilde{M}, \\
  0, & m = 0,
\end{cases}
\end{eqnarray}
where the first condition, i.e., $m = 1, \dots, \tilde{M}$ denotes modulation or active state at the tag, whereas the second condition, i.e., $m = 0$, indicates the inactive state or impedance matching/EH without reflection at the tag. In addition, $Z_m(k)$ is the impedance value, $Z_a$ is the antenna impedance, $\beta$ denotes the tag’s power reflection factor, and $b_m(k)$ is the modulation symbol. Considering a real-valued antenna impedance $Z_a = R_a$, the normalized load impedances can be expressed as
\begin{equation}
z_m = \frac{1 + \Gamma_m}{1 - \Gamma_m} = r_m + jx_m, \quad m = 1, \dots, \tilde{M},
\end{equation}
where $r_m$ and $x_m$ respectively represent the normalized load resistance and reactance.

Modulation strategies span a spectrum, encompassing fundamental methods like on-off keying (OOK) to advanced techniques such as quadrature amplitude modulation (QAM) and phase-shift keying (PSK) \cite{Diluka2022}. The selection of a modulation technique is typically guided by specific application requirements and the targeted data rate.

\begin{figure*}[t]
    \centering 
    \def\svgwidth{480pt} 
    \fontsize{9}{9}\selectfont 
    \graphicspath{{Figures/}}
    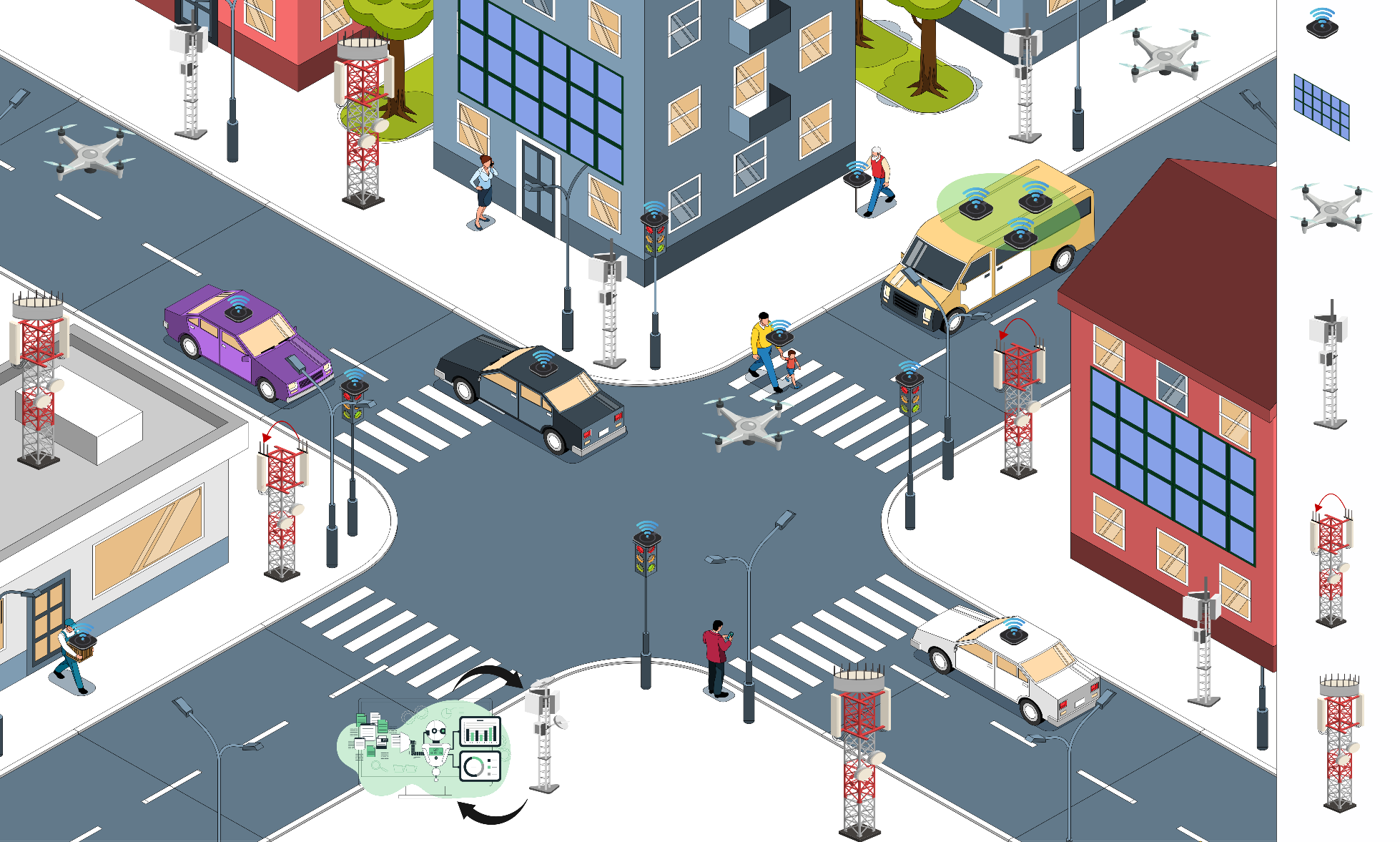  
    \vspace{2mm}
    \caption{Illustration of ISABC configurations, applications, future directions, and open issues.}  \label{fig_ISACApplications}
\end{figure*}

\subsubsection{Backscatter demodulation}
Demodulation involves the reader translating reflected signals from a nearby emitter back to their original data \cite{Diluka2022}. An envelope detector classifies the signal amplitude into `high' or `low' levels to determine the signal's presence or absence from the reader. Due to the randomness in detection, emanating from noise and interference, statistical methods are pivotal. Signal detection uses hypothesis testing with each hypothesis representing unique data probability density functions. The quality of detection is influenced by channel characteristics and the channel state information (CSI). Detection techniques vary: Coherent detection demands full carrier phase and CSI knowledge; Non-coherent detection works without this information; Semi-coherent detection uses limited training symbols; and ML-based detection employs ML algorithms to adapt and improve over time.


\subsubsection{Impedance matching}
To ensure efficient energy transfer from incoming signals, a tag adjusts its impedance to match that of its antenna \cite{Diluka2022}. Perfect matching means there's no reflection and energy is maximally absorbed. If there is a mismatch, the system can become inefficient. This impedance matching ensures that the antenna and tag are working optimally together, avoiding loss of signal or power. When the match is not perfect, additional matching networks are introduced to improve energy transfer.

\subsubsection{Energy harvesting}
In the context of tags, energy harvesting (EH) plays a pivotal role in achieving energy self-sufficiency \cite{Diluka2022}. Tags employ an EH circuit, primarily composed of an antenna and a rectifier, to convert incoming RF signals into direct current (DC). Essential parameters governing the EH process encompass sensitivity and energy transfer efficiency.
For many passive RFID tags, the activation threshold is set around \qty{-20}{\dB m}, representing the minimum power required to activate the EH circuit. The RF-to-DC conversion efficiency, which depends significantly on factors like antenna design, impedance matching, and rectifier efficiency, typically yields real-world power outputs ranging from approximately $\sim$\qty{1}{\uW} to $\sim$\qty{100}{\uW}.

The amount of energy harvested can be modeled as a linear or non-linear process. The linear EH model offers simplicity and it does not capture energy harvesters' nonlinearities due to components such as diodes. At elevated input powers, a saturation emerges, and if the RF is below sensitivity, the output becomes null. Thus, nonlinear models are often more representative \cite{Diluka2022}.

\section{ISABC Applications}
ISABC can facilitate every application that conventional BackCom and ISAC support (Fig.~\ref{fig_BCApplications}).  However,  their challenges are shown in Table \ref{tab:challenges}.

\textbf{Smart agriculture:} Today's precision farming landscape increasingly demands real-time data about soil moisture, temperature, and other vital parameters \cite{Daskalakis2017}. By distributing integrated sensing and backscatter devices across vast farms, farmers gain the ability to glean detailed insights. These insights come without the need for substantial energy consumption or regular maintenance, presenting a significant advantage over traditional methods.

\textbf{Supply chain and logistics:} Ensuring that goods move efficiently and safely is paramount in today's interconnected global economy. Backscatter devices can be embedded within packages or containers, providing real-time insights regarding their location, temperature, humidity, and even tampering. Given their low energy requirements, these devices can operate for extended durations, ensuring continuous monitoring throughout long transit times.

\textbf{Environmental monitoring:} Understanding our environment is vital to combating climate change and ensuring sustainable living. With backscatter devices, researchers can gather accurate data from dense forests, deep oceans, or high altitudes. Monitoring parameters like atmospheric carbon dioxide (CO$_2$) levels, water quality, and wildlife movement becomes seamlessly achievable, providing invaluable data for ecological studies and conservation efforts.

\textbf{Industrial automation:} The Industry $4.0$ revolution is already underway, with factories seeking to become smarter. Backscatter devices can be utilized to monitor machinery health, product quality, and environmental conditions. Instantaneous feedback allows for real-time adjustments, ensuring optimal production outcomes.

\textbf{Health monitoring:} The health sector too stands to gain. Wearable devices equipped with sensors can now constantly monitor vital signs and other crucial health metrics. Through backscatter, these devices can relay data, ensuring continuous monitoring. The minimal energy expenditure makes these devices particularly suitable for long-term patient surveillance.

\textbf{Urban planning:} The vision of smart cities requires a deep understanding of elements such as traffic flows, environmental metrics, and infrastructure utilization. With devices that sense and communicate using backscatter technology, city planners can embed these into urban infrastructures. The longevity of these devices means they can operate for years without necessitating battery replacements, thus providing real-time, uninterrupted data for urban planning and development.

\section{Future Directions and Open Issues}
The convergence of sensing and BackCom is only the tip of the iceberg. As we delve deeper into this century, the merging of various disciplines promises innovations that can redefine our technological landscape. For ISABC systems, several exciting trajectories loom on the horizon (Fig.~\ref{fig_ISACApplications}).

\textbf{Towards more energy-efficient designs:} As miniaturization progresses and the imperative for energy conservation intensifies, we will likely see an influx of designs accentuating ultra-energy efficiency. Ongoing research aims to unveil nano-scale components and circuits with consumption levels in the pico- or femto-watt range. Combined with innovative EH techniques, such as triboelectric nanogenerators or advanced photovoltaic materials, we might be nearing the advent of devices that can operate perpetually, drawing all requisite energy from their surroundings.

\textbf{Massive MIMO and cell-free massive MIMO:}
Massive multiple-input multiple-output (MIMO), through large antenna arrays, offers spatial degrees of freedom (DoF) for multiplexing and diversity gains. In contrast, cell-free architecture enhances coverage probability and harnesses macro-diversity against large-scale fading. ISABC integration with these techniques must take advantage of the favorable properties of massive MIMO and cell-free massive MIMO architectures to boost joint communication and sensing capabilities. 

\textbf{FD ISABC:} By leveraging full-duplex (FD) capabilities, our versatile transmitter can simultaneously delve into EH and initiate active wireless broadcasting. This necessitates the optimal regulation of the hybrid transmitter, particularly given the inherent benefits of full-duplex operations. Future works include refining the optimal power allocation at the BS and considering interference mitigation strategies.

\textbf{UAVs-aided ISABC:} As unmanned aerial vehicles (UAVs) permeate sectors from logistics to agriculture, integrating with ISAC offers transformative potential. ISAC's inclusion reduces the UAV's communication and sensing load, lightening its weight and boosting agility. This integration delivers enhanced performance, surpassing traditional UAV functions. 

\textbf{IRSs-aided ISABC:} The melding of ISABC with intelligent reflecting surfaces (IRSs), which modulates the wireless environment itself, opens up a plethora of opportunities. IRS can be strategically positioned to improve the efficiency of EH, focusing ambient RF signals onto ISABC-equipped devices. Moreover, in active communication scenarios, the IRS can help in sculpting the wireless channels, leading to optimized data transmission and potentially reducing interference

\textbf{Multiple access techniques:} The integration of advanced multiple access techniques like non-orthogonal multiple access (NOMA) or sparse code multiple access (SCMA) with ISABC can significantly augment the simultaneous communication capabilities of these systems. These techniques enable devices to share spectral resources, vital for ISABC systems prioritizing energy efficiency and spectrum optimization.

\textbf{Machine learning in ISABC:} The escalating prowess of ML algorithms signifies the monumental potential for integrated systems. In-device processing can pave the way for smarter data acquisition; sensors might only forward data upon anomaly detection, conserving energy. Beyond that, ML stands poised to refine communication protocols, discerning the ambient RF milieu, ensuring ideal modulation schemes, and curbing interference. This can culminate in comprehensive smart ecosystems that dynamically adapt and respond to varying external factors.

\textbf{Network configurations:} The potential of hybrid transmitters could be notably amplified within a network framework, incorporating multiple tags and a myriad of licensed channels. This structure promises abundant signal resources for energy absorption and an expanded array of spectrum vacancies for concurrent active data communications. Yet, this multifaceted environment amplifies coordination intricacies among the hybrid transmitters, introducing a nuanced equilibrium among EH, backscattering, active broadcasting, and channel selection. 

\begin{figure}[!t]\centering \vspace{-0mm}
\def\svgwidth{210pt} 
\fontsize{8}{8}\selectfont 
\graphicspath{{Figures/}}
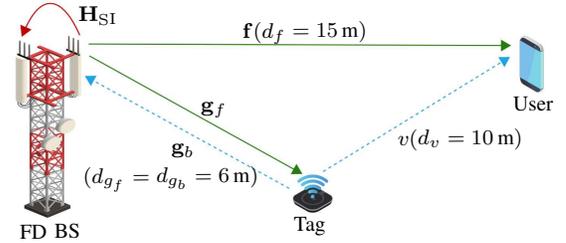 
\vspace{2mm}
\caption{An ISABC system setup.}\vspace{-4mm} \label{fig_SystemModel}
\end{figure}

\section{Numerical Results}
This section presents simulations of both communications (i.e., direct and backscatter) and sensing performance to illustrate the feasibility and performance gains of ISABC. Interested readers can refer to \cite{Diluka2023} for the technical details. We model large-scale fading using the third-generation partnership project (3GPP) urban microcell (UMi) model and the additive white Gaussian noise (AWGN). We consider the conventional ISAC as a comparative benchmark, i.e., the tag is replaced by a conventional radar target with no reflection loss.

\begin{figure}[t]\vspace{-4mm}
\centering
\includegraphics[width=3.5in]{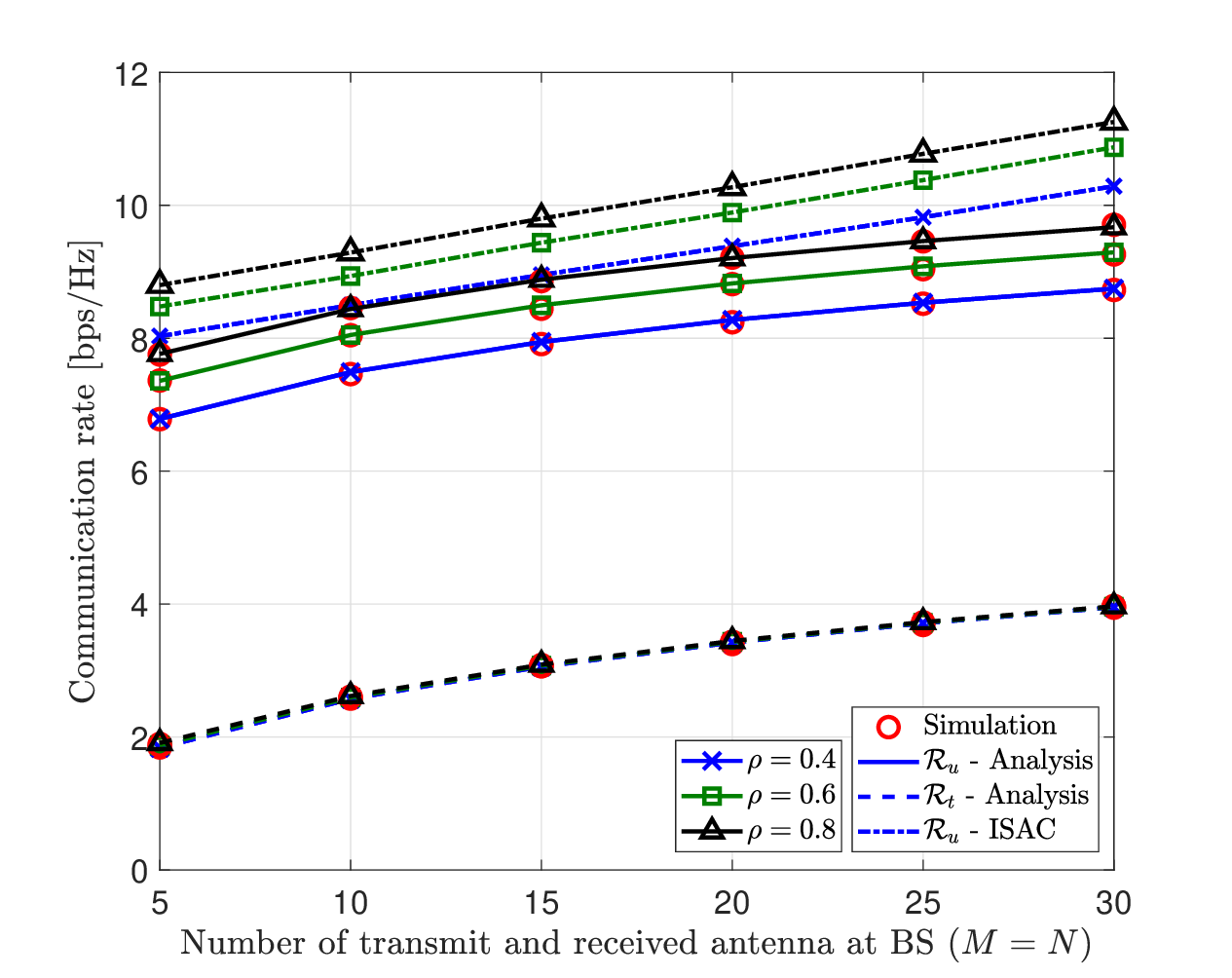}
\caption{Communication rate versus the number of transmit and receive antenna at the BS ($M=N$) for $\rho = \{0.4, 0.6, 0.8\}$.}
\label{fig_CommRate_PtxComp}  \vspace{-4mm}
\end{figure}

Fig. \ref{fig_CommRate_PtxComp} represents the communication rates of both the user and the tag against the number of BS antennas, $M=N$. It is depicted under a BS transmit power of \qty{20}{\dB m}, and for $\rho= \{0.4, 0.6, 0.8\}$, where $\rho$ denotes the power allocation factor of the BS between communication and sensing. The tag rate is independent of $\rho$, while a high $\rho$ contributes to a higher user rate. For instance, with $M=N=20$, a user rate increases by \qty{6.65}{\percent} when $\rho$ changes from \num{0.4} to \num{0.6}. However, due to the absence of tag interference, the ISAC benchmark user rate is always higher than the proposed ISABC user rate. 

\begin{figure}[t]\vspace{-4mm}
\centering
\includegraphics[width=3.5in]{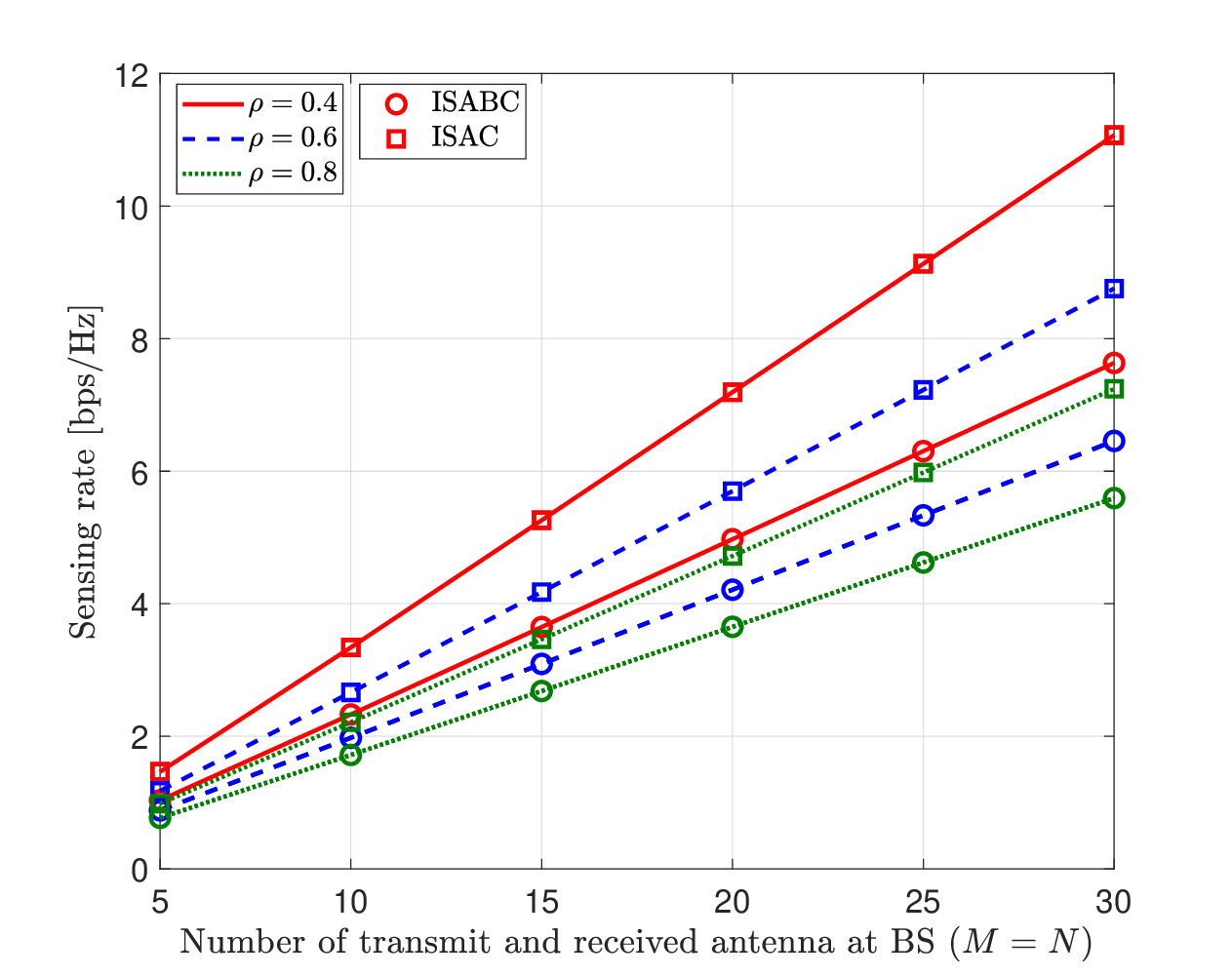}
\caption{Sensing rate versus the number of transmit and receive antenna at the BS ($M=N$) for $\rho = \{0.4, 0.6, 0.8\}$.}
\label{fig_SensRate_PtxComp}  \vspace{-4mm}
\end{figure}

Fig. \ref{fig_SensRate_PtxComp} depicts the BS’s sensing rate versus the number of BS antennas. As expected, increasing the number of BS antennas improves the sensing rate of the BS. Furthermore, low $\rho$ values result in a high sensing rate due to the high power allocation for the sensing waveform. For example, $\rho=0.4$ achieves a \qty{36.16}{\percent} sensing rate gain compared to that of $\rho=0.8$ due to the allocated high power for sensing with $M=N=20$. Moreover, the ISAC system achieves higher sensing rates than the ISABC system since there is no reflection loss at the target. Nevertheless, the proposed system outperforms ISAC. For instance, the ISABC system delivers a \qty{6.7}{\percent} sum rate gain with $M=N=30$ and $\rho= 0.6$.

\section{Conclusion}
The rapidly expanding digital landscape is anchored by the IoT, facilitated majorly by mMTC. However, mMTC faces challenges like energy constraints and effective communication. AmBC offers solutions through its energy efficiency and cost-effectiveness. The merger of ISAC with backscatter evolves into ISABC, enhancing sensory accuracy and communication depth but also adding decoding complexities. The importance of passive tags in ISABC is emphasized, especially in terms of antenna design and EH. In conclusion, while IoT, AmBC, and ISABC herald a future of seamless communication and sensing, hurdles still exist. Future research could improve decoding in ISABC systems and refine EH methods, aiming for an efficient and interconnected future.

\bibliographystyle{IEEEtran}
\bibliography{IEEEabrv,ref}

\end{document}

%% file: Figures/BCApplications.eps_tex
\begingroup%
  \makeatletter%
  \providecommand\color[2][]{%
    \errmessage{(Inkscape) Color is used for the text in Inkscape, but the package 'color.sty' is not loaded}%
    \renewcommand\color[2][]{}%
  }%
  \providecommand\transparent[1]{%
    \errmessage{(Inkscape) Transparency is used (non-zero) for the text in Inkscape, but the package 'transparent.sty' is not loaded}%
    \renewcommand\transparent[1]{}%
  }%
  \providecommand\rotatebox[2]{#2}%
  \newcommand*\fsize{\dimexpr\f@size pt\relax}%
  \newcommand*\lineheight[1]{\fontsize{\fsize}{#1\fsize}\selectfont}%
  \ifx\svgwidth\undefined%
    \setlength{\unitlength}{537.17243958bp}%
    \ifx\svgscale\undefined%
      \relax%
    \else%
      \setlength{\unitlength}{\unitlength * \real{\svgscale}}%
    \fi%
  \else%
    \setlength{\unitlength}{\svgwidth}%
  \fi%
  \global\let\svgwidth\undefined%
  \global\let\svgscale\undefined%
  \makeatother%
  \begin{picture}(1,0.37112533)%
    \lineheight{1}%
    \setlength\tabcolsep{0pt}%
    \put(0,0){\includegraphics[width=\unitlength]{BCApplications.eps}}%
    \put(0.04969417,0.30067904){\rotatebox{32.933147}{\makebox(0,0)[lt]{\lineheight{1.25}\smash{\begin{tabular}[t]{l}Smart Homes\end{tabular}}}}}%
    \put(0.02234471,0.09464843){\rotatebox{-29.830908}{\makebox(0,0)[lt]{\lineheight{1.25}\smash{\begin{tabular}[t]{l}Shopping Centre\end{tabular}}}}}%
    \put(0.8917565,0.3700483){\rotatebox{-35.29031}{\makebox(0,0)[lt]{\lineheight{1.25}\smash{\begin{tabular}[t]{l}Smart Cities\end{tabular}}}}}%
    \put(0.35832015,0.36354656){\makebox(0,0)[lt]{\lineheight{1.25}\smash{\begin{tabular}[t]{l}Logistics\end{tabular}}}}%
    \put(0.63319241,0.36242895){\makebox(0,0)[lt]{\lineheight{1.25}\smash{\begin{tabular}[t]{l}Healthcare\end{tabular}}}}%
    \put(0.63650574,0.00086605){\makebox(0,0)[lt]{\lineheight{1.25}\smash{\begin{tabular}[t]{l}Transportation\end{tabular}}}}%
    \put(0.90057411,0.02254357){\rotatebox{30.6598}{\makebox(0,0)[lt]{\lineheight{1.25}\smash{\begin{tabular}[t]{l}Agriculture\end{tabular}}}}}%
    \put(0.36497115,0.00188883){\makebox(0,0)[lt]{\lineheight{1.25}\smash{\begin{tabular}[t]{l}Industry\end{tabular}}}}%
    \put(0.67858935,0.14103439){\makebox(0,0)[lt]{\lineheight{1.25}\smash{\begin{tabular}[t]{l}AmBC\end{tabular}}}}%
    \put(0.466367,0.14103439){\makebox(0,0)[lt]{\lineheight{1.25}\smash{\begin{tabular}[t]{l}BiBC\end{tabular}}}}%
    \put(0.26252184,0.14382679){\makebox(0,0)[lt]{\lineheight{1.25}\smash{\begin{tabular}[t]{l}MoBC\end{tabular}}}}%
    \put(0.20494995,0.16823323){\makebox(0,0)[lt]{\lineheight{1.25}\smash{\begin{tabular}[t]{l}\tiny{Reader}\end{tabular}}}}%
    \put(0.34177752,0.17381803){\makebox(0,0)[lt]{\lineheight{1.25}\smash{\begin{tabular}[t]{l}\tiny{Tag}\end{tabular}}}}%
    \put(0.4534735,0.16823323){\makebox(0,0)[lt]{\lineheight{1.25}\smash{\begin{tabular}[t]{l}\tiny{Tag}\end{tabular}}}}%
    \put(0.67965781,0.16823323){\makebox(0,0)[lt]{\lineheight{1.25}\smash{\begin{tabular}[t]{l}\tiny{Tag}\end{tabular}}}}%
    \put(0.55120744,0.17940283){\makebox(0,0)[lt]{\lineheight{1.25}\smash{\begin{tabular}[t]{l}\tiny{Reader}\end{tabular}}}}%
    \put(0.77180699,0.16823323){\makebox(0,0)[lt]{\lineheight{1.25}\smash{\begin{tabular}[t]{l}\tiny{Reader}\end{tabular}}}}%
    \put(0.38924831,0.14310164){\makebox(0,0)[lt]{\lineheight{1.25}\smash{\begin{tabular}[t]{l}\tiny{Emitter}\end{tabular}}}}%
    \put(0.60426302,0.16264843){\makebox(0,0)[lt]{\lineheight{1.25}\smash{\begin{tabular}[t]{l}\tiny{RF Source}\end{tabular}}}}%
    \put(0.7243362,0.24921281){\makebox(0,0)[lt]{\lineheight{1.25}\smash{\begin{tabular}[t]{l}\tiny{User}\end{tabular}}}}%
  \end{picture}%
\endgroup%

%% file: Figures/SigProDiagram.eps_tex
\begingroup%
  \makeatletter%
  \providecommand\color[2][]{%
    \errmessage{(Inkscape) Color is used for the text in Inkscape, but the package 'color.sty' is not loaded}%
    \renewcommand\color[2][]{}%
  }%
  \providecommand\transparent[1]{%
    \errmessage{(Inkscape) Transparency is used (non-zero) for the text in Inkscape, but the package 'transparent.sty' is not loaded}%
    \renewcommand\transparent[1]{}%
  }%
  \providecommand\rotatebox[2]{#2}%
  \newcommand*\fsize{\dimexpr\f@size pt\relax}%
  \newcommand*\lineheight[1]{\fontsize{\fsize}{#1\fsize}\selectfont}%
  \ifx\svgwidth\undefined%
    \setlength{\unitlength}{429.54743958bp}%
    \ifx\svgscale\undefined%
      \relax%
    \else%
      \setlength{\unitlength}{\unitlength * \real{\svgscale}}%
    \fi%
  \else%
    \setlength{\unitlength}{\svgwidth}%
  \fi%
  \global\let\svgwidth\undefined%
  \global\let\svgscale\undefined%
  \makeatother%
  \begin{picture}(1,0.64688387)%
    \lineheight{1}%
    \setlength\tabcolsep{0pt}%
    \put(0,0){\includegraphics[width=\unitlength]{SigProDiagram.eps}}%
    \put(0.03897147,0.08781052){\makebox(0,0)[lt]{\lineheight{1.25}\smash{\begin{tabular}[t]{l}Users\\\end{tabular}}}}%
    \put(0.04246353,0.09479452){\makebox(0,0)[lt]{\lineheight{1.25}\smash{\begin{tabular}[t]{l}\\Data\end{tabular}}}}%
    \put(0.11579651,0.08781052){\makebox(0,0)[lt]{\lineheight{1.25}\smash{\begin{tabular}[t]{l}Sensing\end{tabular}}}}%
    \put(0.15486867,0.00805457){\makebox(0,0)[lt]{\lineheight{1.25}\smash{\begin{tabular}[t]{l}(a) FD ISABC BS\end{tabular}}}}%
    \put(0.6736802,0.34835031){\makebox(0,0)[lt]{\lineheight{1.25}\smash{\begin{tabular}[t]{l}(b) ISABC Tag\end{tabular}}}}%
    \put(0.11928856,0.09479452){\makebox(0,0)[lt]{\lineheight{1.25}\smash{\begin{tabular}[t]{l}\\Signal\end{tabular}}}}%
    \put(0.02217074,0.03249858){\makebox(0,0)[lt]{\lineheight{1.25}\smash{\begin{tabular}[t]{l}Transmitter RF Chain\end{tabular}}}}%
    \put(0.2875666,0.03249858){\makebox(0,0)[lt]{\lineheight{1.25}\smash{\begin{tabular}[t]{l}Receiver RF Chain\end{tabular}}}}%
    \put(0.6457442,0.04454522){\makebox(0,0)[lt]{\lineheight{1.25}\smash{\begin{tabular}[t]{l}(c) ISABC User/Reader\end{tabular}}}}%
    \put(0.03198738,0.31479376){\makebox(0,0)[lt]{\lineheight{1.25}\smash{\begin{tabular}[t]{l}Transmitter Signal\end{tabular}}}}%
    \put(0.06341584,0.32177776){\makebox(0,0)[lt]{\lineheight{1.25}\smash{\begin{tabular}[t]{l}\\Processing\end{tabular}}}}%
    \put(0.04418858,0.18019543){\rotatebox{90}{\makebox(0,0)[lt]{\lineheight{1.25}\smash{\begin{tabular}[t]{l}Modulation\end{tabular}}}}}%
    \put(0.12863651,0.17825267){\rotatebox{90}{\makebox(0,0)[lt]{\lineheight{1.25}\smash{\begin{tabular}[t]{l}Multiplexing\end{tabular}}}}}%
    \put(0.17613301,0.17886998){\rotatebox{90}{\makebox(0,0)[lt]{\lineheight{1.25}\smash{\begin{tabular}[t]{l}Interleaving\end{tabular}}}}}%
    \put(0.08609259,0.18019543){\rotatebox{90}{\makebox(0,0)[lt]{\lineheight{1.25}\smash{\begin{tabular}[t]{l}Coding\end{tabular}}}}}%
    \put(0.2924154,0.22552592){\makebox(0,0)[lt]{\lineheight{1.25}\smash{\begin{tabular}[t]{l}Receiver Signal\end{tabular}}}}%
    \put(0.31336771,0.23250992){\makebox(0,0)[lt]{\lineheight{1.25}\smash{\begin{tabular}[t]{l}\\Processing\end{tabular}}}}%
    \put(0.40303236,0.07152459){\rotatebox{90}{\makebox(0,0)[lt]{\lineheight{1.25}\smash{\begin{tabular}[t]{l}Multi-Tag Det-\end{tabular}}}}}%
    \put(0.35699643,0.07346735){\rotatebox{90}{\makebox(0,0)[lt]{\lineheight{1.25}\smash{\begin{tabular}[t]{l}Range/Velocity\end{tabular}}}}}%
    \put(0.29414047,0.07324909){\rotatebox{90}{\makebox(0,0)[lt]{\lineheight{1.25}\smash{\begin{tabular}[t]{l}Com. Data \end{tabular}}}}}%
    \put(0.31160048,0.07346735){\rotatebox{90}{\makebox(0,0)[lt]{\lineheight{1.25}\smash{\begin{tabular}[t]{l}Removing\end{tabular}}}}}%
    \put(0.39954036,0.06957585){\rotatebox{90}{\makebox(0,0)[lt]{\lineheight{1.25}\smash{\begin{tabular}[t]{l}\\-ection/Tracking\end{tabular}}}}}%
    \put(0.03886234,0.39587476){\makebox(0,0)[lt]{\lineheight{1.25}\smash{\begin{tabular}[t]{l}Digital to Analog\end{tabular}}}}%
    \put(0.03908059,0.47968396){\makebox(0,0)[lt]{\lineheight{1.25}\smash{\begin{tabular}[t]{l}RF Upconversion\end{tabular}}}}%
    \put(0.03897147,0.56272922){\makebox(0,0)[lt]{\lineheight{1.25}\smash{\begin{tabular}[t]{l}Power Amplifier\end{tabular}}}}%
    \put(0.82782377,0.22629312){\makebox(0,0)[lt]{\lineheight{1.25}\smash{\begin{tabular}[t]{l}Signal Processing\end{tabular}}}}%
    \put(0.87754161,0.22891935){\makebox(0,0)[lt]{\lineheight{1.25}\smash{\begin{tabular}[t]{l}\\for\end{tabular}}}}%
    \put(0.85309725,0.20796735){\makebox(0,0)[lt]{\lineheight{1.25}\smash{\begin{tabular}[t]{l}\\User Data\end{tabular}}}}%
    \put(0.5135388,0.16343627){\makebox(0,0)[lt]{\lineheight{1.25}\smash{\begin{tabular}[t]{l}Signal Processing\end{tabular}}}}%
    \put(0.56325664,0.16606249){\makebox(0,0)[lt]{\lineheight{1.25}\smash{\begin{tabular}[t]{l}\\for\end{tabular}}}}%
    \put(0.53182818,0.14511049){\makebox(0,0)[lt]{\lineheight{1.25}\smash{\begin{tabular}[t]{l}\\Backscatter\end{tabular}}}}%
    \put(0.56325663,0.09971448){\makebox(0,0)[lt]{\lineheight{1.25}\smash{\begin{tabular}[t]{l}Data\end{tabular}}}}%
    \put(0.29028976,0.36444631){\makebox(0,0)[lt]{\lineheight{1.25}\smash{\begin{tabular}[t]{l}Analog to Digital\end{tabular}}}}%
    \put(0.28003187,0.42730324){\makebox(0,0)[lt]{\lineheight{1.25}\smash{\begin{tabular}[t]{l}RF Downconversion\end{tabular}}}}%
    \put(0.31135119,0.56971332){\makebox(0,0)[lt]{\lineheight{1.25}\smash{\begin{tabular}[t]{l}Low Noise \\\end{tabular}}}}%
    \put(0.3183353,0.57560609){\makebox(0,0)[lt]{\lineheight{1.25}\smash{\begin{tabular}[t]{l} \\Amplifier\end{tabular}}}}%
    \put(0.57992361,0.453168){\makebox(0,0)[lt]{\lineheight{1.25}\smash{\begin{tabular}[t]{l}Receiver\end{tabular}}}}%
    \put(0.70198309,0.46041366){\makebox(0,0)[lt]{\lineheight{1.25}\smash{\begin{tabular}[t]{l}\\Controller \\\end{tabular}}}}%
    \put(0.71944335,0.46041366){\makebox(0,0)[lt]{\lineheight{1.25}\smash{\begin{tabular}[t]{l}Micro\\\\\end{tabular}}}}%
    \put(0.57992361,0.50554874){\makebox(0,0)[lt]{\lineheight{1.25}\smash{\begin{tabular}[t]{l}EH Unit\end{tabular}}}}%
    \put(0.82785907,0.50554874){\makebox(0,0)[lt]{\lineheight{1.25}\smash{\begin{tabular}[t]{l}Data Storage\end{tabular}}}}%
    \put(0.82785907,0.44618394){\makebox(0,0)[lt]{\lineheight{1.25}\smash{\begin{tabular}[t]{l}Sensing Unit\end{tabular}}}}%
    \put(0.85230343,0.4077713){\makebox(0,0)[lt]{\lineheight{1.25}\smash{\begin{tabular}[t]{l}Other\end{tabular}}}}%
    \put(0.83135112,0.3903113){\makebox(0,0)[lt]{\lineheight{1.25}\smash{\begin{tabular}[t]{l}Components\end{tabular}}}}%
    \put(0.57293953,0.40078728){\makebox(0,0)[lt]{\lineheight{1.25}\smash{\begin{tabular}[t]{l}Modulator\end{tabular}}}}%
    \put(0.53499895,0.2372964){\makebox(0,0)[lt]{\lineheight{1.25}\smash{\begin{tabular}[t]{l}RF Down-\\\end{tabular}}}}%
    \put(0.5315069,0.2407884){\makebox(0,0)[lt]{\lineheight{1.25}\smash{\begin{tabular}[t]{l}\\-conversion\\\end{tabular}}}}%
    \put(0.68515702,0.2372964){\makebox(0,0)[lt]{\lineheight{1.25}\smash{\begin{tabular}[t]{l}Analog to\end{tabular}}}}%
    \put(0.69563317,0.2407884){\makebox(0,0)[lt]{\lineheight{1.25}\smash{\begin{tabular}[t]{l}\\Digital\\\end{tabular}}}}%
    \put(0.72706161,0.11158277){\makebox(0,0)[lt]{\lineheight{1.25}\smash{\begin{tabular}[t]{l}SIC\end{tabular}}}}%
    \put(0.69865329,0.55792949){\makebox(0,0)[lt]{\lineheight{1.25}\smash{\begin{tabular}[t]{l}Energy Storage\end{tabular}}}}%
    \put(0.31135119,0.49987232){\makebox(0,0)[lt]{\lineheight{1.25}\smash{\begin{tabular}[t]{l}Analog SI \\\end{tabular}}}}%
    \put(0.30436709,0.50511027){\makebox(0,0)[lt]{\lineheight{1.25}\smash{\begin{tabular}[t]{l}\\Cancellation\end{tabular}}}}%
    \put(0.17972096,0.62336709){\makebox(0,0)[lt]{\lineheight{1.25}\smash{\begin{tabular}[t]{l}Antenna SI \\\end{tabular}}}}%
    \put(0.17622891,0.63086242){\makebox(0,0)[lt]{\lineheight{1.25}\smash{\begin{tabular}[t]{l}\\Cancellation\end{tabular}}}}%
    \put(0.31484324,0.30431763){\makebox(0,0)[lt]{\lineheight{1.25}\smash{\begin{tabular}[t]{l}Digital SI \\\end{tabular}}}}%
    \put(0.30436709,0.30955558){\makebox(0,0)[lt]{\lineheight{1.25}\smash{\begin{tabular}[t]{l}\\Cancellation\end{tabular}}}}%
  \end{picture}%
\endgroup%

%% file: Figures/ISABCConfigurations1.eps_tex
\begingroup%
  \makeatletter%
  \providecommand\color[2][]{%
    \errmessage{(Inkscape) Color is used for the text in Inkscape, but the package 'color.sty' is not loaded}%
    \renewcommand\color[2][]{}%
  }%
  \providecommand\transparent[1]{%
    \errmessage{(Inkscape) Transparency is used (non-zero) for the text in Inkscape, but the package 'transparent.sty' is not loaded}%
    \renewcommand\transparent[1]{}%
  }%
  \providecommand\rotatebox[2]{#2}%
  \newcommand*\fsize{\dimexpr\f@size pt\relax}%
  \newcommand*\lineheight[1]{\fontsize{\fsize}{#1\fsize}\selectfont}%
  \ifx\svgwidth\undefined%
    \setlength{\unitlength}{939.28015137bp}%
    \ifx\svgscale\undefined%
      \relax%
    \else%
      \setlength{\unitlength}{\unitlength * \real{\svgscale}}%
    \fi%
  \else%
    \setlength{\unitlength}{\svgwidth}%
  \fi%
  \global\let\svgwidth\undefined%
  \global\let\svgscale\undefined%
  \makeatother%
  \begin{picture}(1,0.6006615)%
    \lineheight{1}%
    \setlength\tabcolsep{0pt}%
    \put(0,0){\includegraphics[width=\unitlength]{ISABCConfigurations1.eps}}%
    \put(0.92979012,0.5565181){\makebox(0,0)[lt]{\lineheight{1.25}\smash{\begin{tabular}[t]{l}Tag\end{tabular}}}}%
    \put(0.92979012,0.48465455){\makebox(0,0)[lt]{\lineheight{1.25}\smash{\begin{tabular}[t]{l}IRS\end{tabular}}}}%
    \put(0.93298405,0.41598496){\makebox(0,0)[lt]{\lineheight{1.25}\smash{\begin{tabular}[t]{l}UAV\end{tabular}}}}%
    \put(0.93777496,0.27385484){\makebox(0,0)[lt]{\lineheight{1.25}\smash{\begin{tabular}[t]{l}AP\end{tabular}}}}%
    \put(0.92499921,0.13332169){\makebox(0,0)[lt]{\lineheight{1.25}\smash{\begin{tabular}[t]{l}FD BS\end{tabular}}}}%
    \put(0.27719764,0.01138657){\makebox(0,0)[lt]{\lineheight{1.25}\smash{\begin{tabular}[t]{l}ML\end{tabular}}}}%
    \put(0.91701438,0.00237034){\makebox(0,0)[lt]{\lineheight{1.25}\smash{\begin{tabular}[t]{l}MIMO BS\end{tabular}}}}%
  \end{picture}%
\endgroup%

%% file: Figures/SystemModel.eps_tex
\begingroup%
  \makeatletter%
  \providecommand\color[2][]{%
    \errmessage{(Inkscape) Color is used for the text in Inkscape, but the package 'color.sty' is not loaded}%
    \renewcommand\color[2][]{}%
  }%
  \providecommand\transparent[1]{%
    \errmessage{(Inkscape) Transparency is used (non-zero) for the text in Inkscape, but the package 'transparent.sty' is not loaded}%
    \renewcommand\transparent[1]{}%
  }%
  \providecommand\rotatebox[2]{#2}%
  \newcommand*\fsize{\dimexpr\f@size pt\relax}%
  \newcommand*\lineheight[1]{\fontsize{\fsize}{#1\fsize}\selectfont}%
  \ifx\svgwidth\undefined%
    \setlength{\unitlength}{338.00769043bp}%
    \ifx\svgscale\undefined%
      \relax%
    \else%
      \setlength{\unitlength}{\unitlength * \real{\svgscale}}%
    \fi%
  \else%
    \setlength{\unitlength}{\svgwidth}%
  \fi%
  \global\let\svgwidth\undefined%
  \global\let\svgscale\undefined%
  \makeatother%
  \begin{picture}(1,0.4387681)%
    \lineheight{1}%
    \setlength\tabcolsep{0pt}%
    \put(0,0){\includegraphics[width=\unitlength]{SystemModel.eps}}%
    \put(0.02952702,0.00774468){\color[rgb]{0,0,0}\rotatebox{0.53733522}{\makebox(0,0)[lt]{\lineheight{1.25}\smash{\begin{tabular}[t]{l}FD BS\end{tabular}}}}}%
    \put(0.13461647,0.39982744){\color[rgb]{0,0,0}\rotatebox{0.53733522}{\makebox(0,0)[lt]{\lineheight{1.25}\smash{\begin{tabular}[t]{l}$\mathbf{H}_{\rm{SI}}$\end{tabular}}}}}%
    \put(0.43028805,0.37001564){\color[rgb]{0,0,0}\rotatebox{0.53733522}{\makebox(0,0)[lt]{\lineheight{1.25}\smash{\begin{tabular}[t]{l}$\mathbf{f} (d_f=15\,\text{m})$\end{tabular}}}}}%
    \put(0.14309153,0.10800831){\color[rgb]{0,0,0}\rotatebox{0.53733522}{\makebox(0,0)[lt]{\lineheight{1.25}\smash{\begin{tabular}[t]{l}$(d_{g_f}=d_{g_b}=6\,\text{m})$\end{tabular}}}}}%
    \put(0.35518584,0.23742883){\color[rgb]{0,0,0}\rotatebox{0.53733522}{\makebox(0,0)[lt]{\lineheight{1.25}\smash{\begin{tabular}[t]{l}$\mathbf{g}_{f}$\end{tabular}}}}}%
    \put(0.70957967,0.17704154){\color[rgb]{0,0,0}\rotatebox{0.53733522}{\makebox(0,0)[lt]{\lineheight{1.25}\smash{\begin{tabular}[t]{l}$v (d_v=10\,\text{m})$\end{tabular}}}}}%
    \put(0.52322883,0.02216808){\color[rgb]{0,0,0}\rotatebox{0.53733522}{\makebox(0,0)[lt]{\lineheight{1.25}\smash{\begin{tabular}[t]{l}Tag\end{tabular}}}}}%
    \put(0.91804387,0.2389121){\color[rgb]{0,0,0}\rotatebox{0.53733522}{\makebox(0,0)[lt]{\lineheight{1.25}\smash{\begin{tabular}[t]{l}User\end{tabular}}}}}%
    \put(0.30193196,0.16198723){\color[rgb]{0,0,0}\rotatebox{0.53733522}{\makebox(0,0)[lt]{\lineheight{1.25}\smash{\begin{tabular}[t]{l}$\mathbf{g}_{b}$\end{tabular}}}}}%
  \end{picture}%
\endgroup%